# Local transformation of the Electronic Structure and Generation of Free Carriers in Cuprates and Ferropnictides under Heterovalent and Isovalent Doping

Kirill Mitsen[1] and Olga Ivanenko

Lebedev Physical Institute, Leninsky pr., 53 Moscow, 119991, Russia

**Abstract**

We have previously shown that most of the anomalies in the superconducting characteristics of cuprates and ferropnictides observed at dopant concentrations within the superconducting dome, as well as the position of the domes in the phase diagrams, do not require knowledge of the details of their electronic structure for explanation, but can be understood and calculated with high accuracy within the framework of a simple model describing the cluster structure of the superconducting phase. This fact suggests a change in the paradigm that forms our understanding of HTSC. In this paper, we propose a unified view on the transformation of the electronic structure of cuprates and ferropnictides upon heterovalent and isovalent doping, based on the assumption of self-localization of doped carriers. In this representation, in undoped cuprates and ferropnictides, which initially have different electronic structures (Mott insulator and semimetal), local doping forms percolation clusters with the same electronic structure of a self-doped excitonic insulator where a specific mechanism of superconducting pairing is implemented, which is genetically inherent in such a system. The proposed model includes a mechanism for generating additional free carriers under heterovalent and isovalent doping and makes it possible to predict their sign, which, in the general case, does not coincide with the sign of doped carriers.

**Introduction**

Elucidation of the mechanism of high-temperature superconductivity still remains a topical issue of condensed matter physics. Despite many years of efforts by theorists and an enormous volume of accumulated experimental knowledge of these materials, many of their features have eluded satisfactory explanation. At the same time, we believe that based on the already established experimental facts, it is possible to identify key aspects that open the way to solving the problem of HTSC.

To date, a list of high-temperature superconductor compounds includes mainly representatives of two classes: cuprates and ferropnictides. In an undoped (stoichiometric) state, these compounds (except LiFeAs) are not superconductors. A superconductivity in them emerges only as a result of heterovalent or isovalent doping, i.e., at a partial substitution of atoms of an element with atoms of another element with a different or the same valence, respectively.

---

[1] Correspondence to [mitsen@lebedev.ru]

Under heterovalent doping, both cuprates and ferropnictides experience, in the general case, the same sequence of transitions from an antiferromagnetic to a superconducting state and then to a normal metal, i.e., superconductivity takes place in only a limited range of doping. It is generally accepted that heterovalent doping of cuprates and ferropnictides introduces additional charge carriers into their basal $CuO_2$ and FeAs planes, which, due to some causes, leads in a certain range of concentrations to the initiation of a superconducting pairing in this compound at $T < T_c$. Herewith the Tc(x) dependence curve has a dome shape in both cuprates and ferropnictides, which suggests the similarity of the mechanisms leading, with increasing doping, to the transition of these compounds to a state where the specific mechanism of superconducting pairing can be implemented.

It is surprising, however, that the same dome-shaped Tc(x) dependence is also observed for ferropnictides with isovalent substitution of ions in the basal plane, for example, in $BaFe_2(As_{1-x}P_x)_2$ and $Ba(Fe_{1-x}Ru_x)_2As_2$, and, moreover, as shown by experiment, such doping, as in the case of heterovalent substitution, is accompanied by a change in the concentration of carriers [1, 2].

All these facts make one look closer at the doping process leading to the formation of superconducting phase in initially non-superconducting parent compounds of cuprates and ferropnictides.

Until now, the predominant majority of works have considered doping as a process of a spatially homogeneous change of electronic structure. This approach, however, fails to agree with many experiments where spatial variations of local charge density and superconducting order parameter are observed in doped specimens [3–6]. What is more, many experiments demonstrate that doped carriers in both cuprates and ferropnictides are strictly localized in the nearest vicinity of the dopant [7–11].

Previously, we proposed a unified mechanism of local iso- and heterovalent (hole and electron) doping, which is implemented in superconducting cuprates and ferropnictides [12, 13]. The proposed mechanism assumes the local nature of the transformation of their band structure under doping, which leads to the formation of inhomogeneous cluster structure of the superconducting phase [13], the parameters of which are determined by the concentration of the dopant and crystal structure of the undoped compound.

We have shown that basing only on the knowledge of the crystal structure and type of dopant, it is possible: 1) to accurately determine the positions of superconducting domes in the phase diagrams of HTSC compounds [12], 2) to explain the nature and the positions of sharp peaks in the London penetration depth as well as the positions of anomalies in the anisotropy of resistance depending on the level heterovalent or isovalent doping [13], 3) to understand the nature of "magic" concentration values corresponding to abrupt changes in superconducting properties (1/8 anomaly, etc.) [12].

In other words, we have demonstrated that all the above experimentally observed anomalies of the superconducting characteristics depending on doping can be understood and accurately calculated within the framework of unified model describing the cluster structure of the superconducting phase. It is importantly, the proposed interpretation does not require the introduction of additional theoretical concepts to explain observed anomalies. The possibility of obtaining this information without using any data on the features of their electronic structure can be considered as an indication of the existence of some general and rather crude mechanism that operates regardless of the details of the electronic structure of these compounds and provides superconducting pairing.

In this work, we propose a unified view on the transformation of the electronic structure of cuprates and ferropnictides upon heterovalent and isovalent doping. According to our idea, in undoped cuprates and ferropnictides, which initially have different electronic structures (Mott insulator and semimetal), local doping forms percolation clusters with the same electronic structure of a self-doped excitonic insulator. The proposed model also includes the specific mechanism for generating of extra free carriers under heterovalent and isovalent doping and makes it possible to predict their sign, which, in the general case, does not coincide with the sign of doped carriers. It is also assumed that a specific mechanism of superconducting pairing is realized in the formed clusters, which is genetically inherent in such a system and is due to the interaction of band electrons with excitonic states".

**Heterovalent doping and trionic complexes**

According to [12], in the case of heterovalent doping of cuprates and ferropnictides, the local nature of the transformation of the electronic structure is due to the self-localization of doped carriers as a result of the formation of trion complexes by them. Such a complex binds the doped carrier with charge-transfer excitons (CT-excitons), which are formed in the basal plane under the influence of this very carrier. Note that self-localization takes place as the temperature decreases below a certain localization temperature of doped carriers - $T_{\text{loc}}$.

The possibility for the existence of CT excitons in cuprates and ferropnictides follows from the fact that both classes of compounds *are characterized by a low concentration of charge carriers*, $n<10^{22}$ cm$^{-3}$, which corresponds to a mean distance between carriers, $r_s >0.4$ nm, and exceeds the distance between the anion and cation. This means that the interaction inside the cell is essentially unshielded, which is what enables the existence of well-defined CT excitons [14].

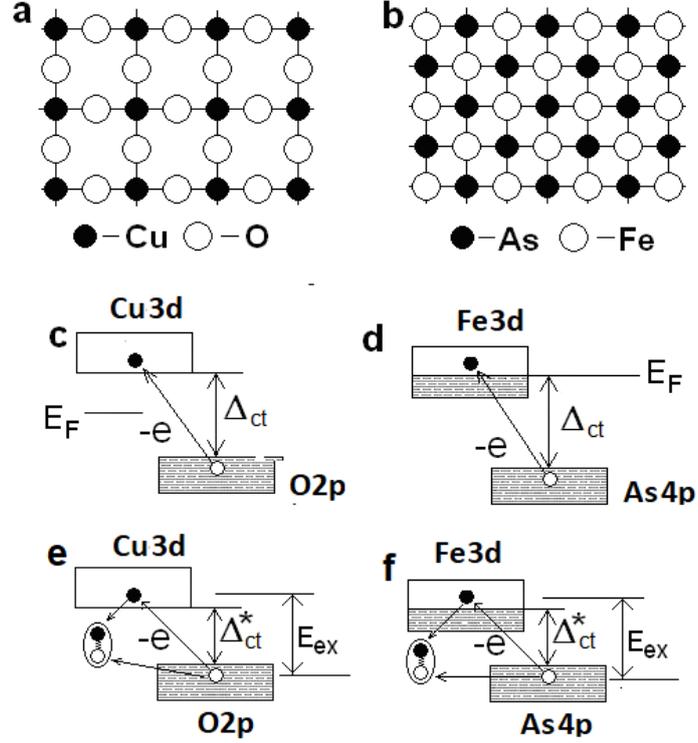

FIG. 1. (a,b) Arrangement of the projections of anions and cations in the basal planes of cuprates (a) and ferropnictides (b); (c,d), the band structures of undoped cuprates and ferropnictides; energy Δ required for interband transition related to the transfer of an electron from an oxygen ion to a copper ion (in cuprates), or to the transfer of a hole from an iron ion to an arsenium ion (in ferropnictides), is approximately the same in both cases and makes ~2 eV; (e,f), the value of $\Delta_{ct}$ can be decreased under the impact of a doped charge localized near the plaquette to a value of $\Delta_{ct}^*$ smaller than CT-exciton bond energy $E_{ex}$. This will lead to a transition to an exciton-bound state of an electron (hole) on the central ion and a hole (electron) on the orbital of the surrounding ions (an intracrystalline analog of the hydrogen atom).

Figure 1 a,b shows the arrangement of anions and cations (their projections, to be more exact) in the basal planes of cuprates and ferropnictides. In cuprates, Cu and O ions lie in one plane; in ferropnictides, Fe ions are in the plane and As ions are at the vertices of the regular tetrahedra such that their projections form a square sublattice in the basal plane. For simplicity of consideration but without the loss of generality, we will consider the FeAs lattice (Fig. 1 b) as a plane lattice and will replace several bands in the vicinity of Fermi energy by one band (Fig. 1 d,f).

The lattices in Fig. 1 a,b can be considered as a net of hydrogen-like plaquettes $CuO_4$ (in cuprates) and $AsFe_4$ or $FeAs_4$ (in ferropnictides). The meaning of singling these plaquettes out is that the energy required for electron (hole) transfer between the central ion and four adjacent ions is approximately the same and equals $\Delta_{ct} \approx 1.5$–$2$ eV [15,16].

This energy, however, can be decreased to a value of $\Delta_{ct}^*$ smaller than CT exciton bond energy $E_{ex}$, which is estimated to equal ~1 eV [14]. That is to say, a transition to the exciton-bound state of an electron (hole) on the central ion and of a hole (electron) on the orbital of the adjacent ions will occur in this plaquette. We will call such plaquettes CT plaquettes.

The value of $\Delta_{ct}$ can be reduced to $\Delta_{ct}^*$, for example, by placing a charge of a certain value on one of the plaquette ions or nearby it . Indeed, within the framework of a simple ionic model, the value of $\Delta_{ct}$ is determined by the following relationship [17]:

$$\Delta_{ct} = E_{cat} - E_{an} = |\Delta E_M| + A - I. \qquad (1)$$

Here, $E_{cat}$ is the minimal energy of an electron transferred to a cation; $E_{an}$, the maximal energy of an electron on an anion; $\Delta E_M$, a change of Madelung energy at a per unit change of cation and anion charge states; $A$, electronic affinity energy of the anion; $I$, corresponding ionization potential of the cation. Taking into account that the value of $\Delta_{ct} \sim 2$ eV, and the Cu and Fe ionization potential $I \sim 20$ eV, the rather slim balance in eq. (1) can be easily changed to either side by changing $|\Delta E_M|$. This is facilitated by the high degree of ionicity of cuprates and ferropnictides, which causes a large contribution of the Madelung energy $E_M$ to the electronic structure of the base planes, which makes it possible to change it locally by introducing localized charges into the crystal.

Obviously, by placing near one of the ions forming the plaquette some additional charge of a certain sign, one can change $|\Delta E_M|$ and, thus, to decrease $\Delta_{ct}$ for this plaquette. If this reduction is sufficient to fulfill the condition $\Delta_{ct}^* < E_{ex}$, then this will lead to the creation of CT plaquette.

The $\Delta_{ct}$ can also be reduced to the required value by placing a charge of a certain magnitude and sign directly on one of the plaquette ions. At the same time, it is important to note that adding a localized negative charge to the cation, as well as a localized positive charge to the anion, will lead to the same result - a local decrease in $|\Delta E_M|$.

Consider the mechanism of formation of CT plaquettes in real structures (Fig. 2). In the cases when dopants are outside the basal plane they can induce the needed charges on ions near plaquettes. For example, in oxygen doped cuprates oxygen dopants outside the $CuO_2$ planes (interstitial oxygens, as in $LaCuO_{4+d}$, or oxygen atoms, occupying vacant sites, as in $YBa_2Cu_3O_{6+d}$ and $Bi_2Sr_2Ca_nCu_{n+1}O_x$) do not supply the holes into the $CuO_2$ planes, but induce the excess positive charges $\sim|e|/4$ on the nearest apical oxygen ions. Each that charge gives rise to one CT plaquette in the $CuO_2$ plane similar to how it takes place in $YBa_2Cu_3O_{6+d}$ (Fig. 2a).

In cuprates and ferropnictides, where a doped carrier (electron, hole) enters the basal plane, it is initially distributed in accordance with the symmetry of the dopant environment on the group orbital of the 4 nearest cations or anions, which is equivalent to the appearance of an excess charge $\sim|e|/4|$ of corresponding sign on them (Fig. 2 b-f). The appearance of such a charge of the corresponding sign on the external or internal ion of the plaquette reduces by ~ 1.5 eV the value of $\Delta_{ct}$ for the transfer of a hole (electron) between the central and external ions of $CuO_4$ or $FeAs_4$ ($AsFe_4$) plaquettes. This decrease in $\Delta_{ct}$ to $\Delta_{ct}^*$, in our opinion, is sufficient for the production of CT exciton in the corresponding plaquette.

Thus, each doped carrier distributed in the group orbital of four neighboring ions generates, in accordance with the symmetry of the environment, several CT plaquettes in the basal plane around the projection of the dopant, which form a trion complex and thus limit the spread of the doped carrier. It is important to note that in the cases when doped carriers are localized outside the $CuO_2$ plane (as, e.g., in $YBa_2Cu_3O_{6+d}$ and $Bi_2Sr_2Ca_nCu_{n+1}O_x$, etc.), it is possible to achieve such a level of doping when the CT plaquettes occupy the entire $CuO_2$ plane.

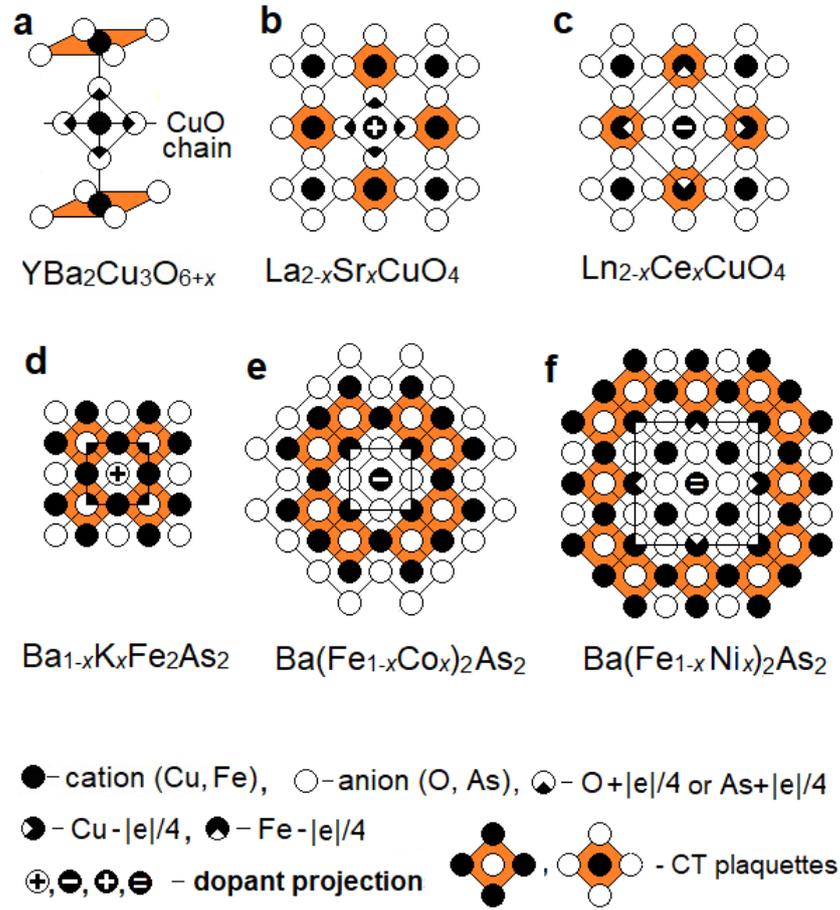

FIG. 2. Formation of trion complexes, consisting of a localized doped carrier and a number of plaquettes [12]. (a) in $YBa_2Cu_3O_{7-x}$ the excess charges are located on apical oxygen ions and two $CuO_4$ plaquettes are formed in $CuO_2$ planes; (b-d) doped carriers are located in basal planes. Ions onto which additional charge $q^*\approx|e|/4$ is sufficient to form a CT plaquette are marked by black or white sectors; (b,c) in $La_{2-x}Sr_xCuO_4$ and $Ln_{2-x}Ce_xCuO_4$ four $CuO_4$ plaquettes are formed in $CuO_2$ planes, (d-f) 4, 8 and 12 $AsFe_4$ plaquettes are formed in FeAs plane in $Ba_{1-x}K_xFe_2As_2$, $Ba(Fe_{1-x}Co_x)_2As_2$ and $Ba(Fe_{1-x}Ni_x)_2As_2$, correspondingly. CT plaquettes are highlighted in brown. For the geometry of trion complexes in other HTSC compounds, see [12].

Note that if we place two charges $\sim|e/4|$ (for example, from two doped carriers), then the gap $\Delta_{ct}$ will disappear, and this will lead to the spread of the doped carrier beyond the first group orbital. In this case, these charges will be distributed over the group orbital of 8 ions following the nearest ones and induce on them the charges $\sim|e/4|$, sufficient for the formation of 8 CT plaquettes. Such a situation, in particular, takes place in $Ba(Fe_{1-x}Ni_x)_2As_2$, where the Ni ion dopes two electrons. Therefore, despite the close relationship between $Ba(Fe_{1-x}Ni_x)_2As_2$ and $Ba(Fe_{1-x}Co_x)_2As_2$, the geometry of their trion complexes is significantly different (Fig. 2 e, f), which causes the difference in the phase diagrams [13].

This, in our opinion, is the essence of heterovalent doping, which is accompanied with the formation of trion complexes. The geometry of such a complex can be determined for each compound, proceeding only from the knowledge of dopant position and surrounding symmetry [12,13]. Figure 2 shows examples of trion complexes in some cuprates and ferropnictides.

The interaction between these trion complexes corresponds to repulsion, which promotes the ordering of dopants and related trions into square lattices with the parameter $l$ depending on the concentration. As the concentration increases, CT plaquettes in neighboring trion complexes begin to bind into a network through 1 or 2 common ions, which is accompanied by the formation of clusters of CT plaquettes characterized by a certain value of $l$.

In a certain range of concentrations, individual clusters of CT plaquettes with the same parameter $l$ form a percolation CT cluster, and it is this cluster that we identify with the HTSC phase [12]. In such a cluster, two-particle transitions back and forth between two single-particle states (p-electron + d-hole) on the one hand and an excitonic two-particle state (d-electron + p-hole) on the other side become possible. Therefore, the state of electron in the CT cluster can be considered as a superposition of the band and exciton states.

Knowing the geometry of trionic complexes (Fig. 2), one can determine the ranges of dopant concentrations corresponding to the existence of a percolation cluster of CT plaquettes. It is easy to check that the ranges of dopant concentrations determined in this way coincide with a high accuracy with the positions of the superconducting domes in the phase diagrams of all the studied compounds. Moreover, this approach made it possible to explain the nature of "magic" values of concentrations corresponding to a sharp change in the superconducting characteristics of a number of compounds with a change in the doping level [12, 13].

**Formation of CT clusters under isovalent doping**

In the case of isovalent doping in ferropnictides, the role of an additional charge that affects the value of $\Delta_{ct}$ in the surrounding plaquettes is played by a change in the electron density near As anions or Fe cations due to the difference in the ionic radii of the dopant and the matrix ion. Thus, in BaFe$_2$(As$_{1-x}$P$_x$)$_2$, the replacement of the As$^{3-}$ ion by a P$^{3-}$ ion with a smaller ionic radius is equivalent to a decrease in the negative charge near four surrounding Fe ions, which reduces the gap $\Delta_{ct}$ for the transfer of an electron to these Fe ions from the nearest As ions to the value $\Delta_{ct}^*$ (Fig. 3a). We will mark such Fe ions as Fe'. Since such a decrease in the gap $\Delta_{ct}$ due to the difference in ionic radii is expected to be less than with heterovalent doping, it is possible to place two P ions next to the Fe ion in order to further reduce the gap to a value of $0<\Delta_{ct}^{**}<\Delta_{ct}^*$. Such Fe ions, whose neighbors are two P ions, we will denote Fe".

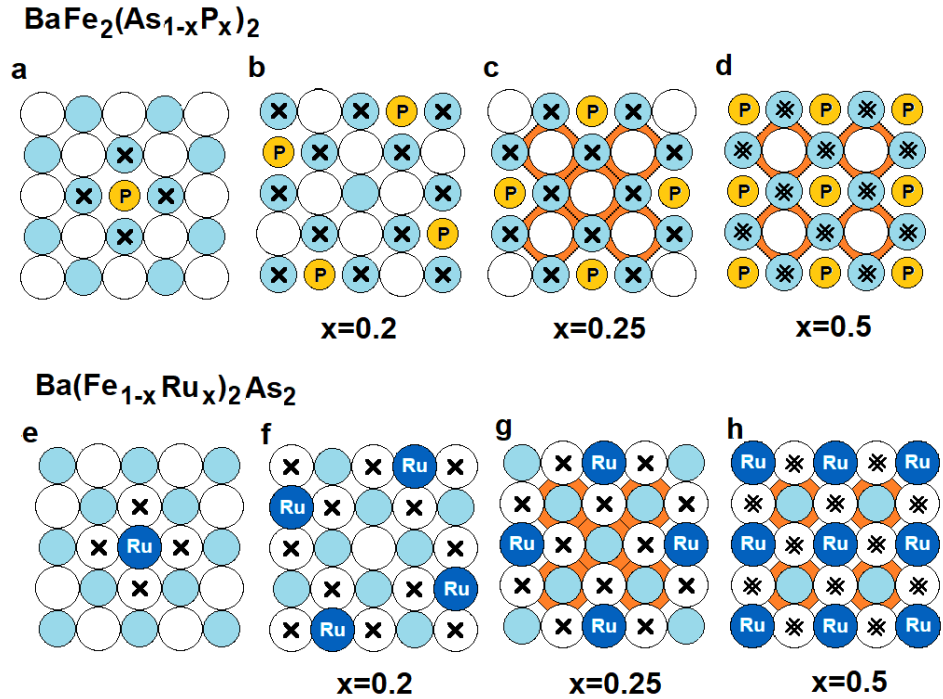

FIG. 3. Formation of CT clusters in Ba122 ferropnictides with isovalent substitution (P, Ru).
(a), substitution of an As ion with a P ion of smaller radius lowers the electron energy at four neighboring Fe' ions; (b), when P ions are ordered into the √5x√5 lattice (x=0.2), AsFe'$_4$ plaquettes are absent; (c), at $x>0.2$, the formation of CT clusters from AsFe'$_4$ plaquettes is possible; (d), at $x>0.3$, the existence of CT clusters of AsFe"$_4$ plaquettes is possible; (e-h), substitution of an Fe ion with a larger Ru ion increases the electron energy on four adjacent As ions. The sequence of transitions from one cluster structure with FeAs'$_4$ plaquettes to another with FeAs'$_4$ plaquettes is similar to Fig.(a-d).

For the formation of AsFe'$_4$ (or AsFe"$_4$) CT plaquettes (shaded in brown), it is necessary that the As ion be surrounded by four Fe' (or Fe") ions. Therefore, as can be seen from Fig. 3b, with an ordered arrangement of P ions with $x=0.2$, CT plaquettes are not formed, and for the formation of a percolation cluster from AsFe'$_4$ or AsFe"$_4$ CT plaquettes, it is necessary that the phosphorus concentration $x$ exceed 0.2. At the same time, as can be seen from Fig. 3 c,d, the formation of large CT clusters is possible with an ordered arrangement of dopants at concentrations $x=0.25$ and 0.5, when the dopants are ordered into lattices with a symmetry that retains the main symmetry elements of the matrix crystal (the Curie principle) [13]. This is confirmed by the appearance at these $x$ values of sharp minima in the dependence of the London penetration depth on doping [18,19]. When the concentration deviates from these special values, the sizes of clusters with this type of ordering rapidly decrease, since this order does not meet the average concentration for homogeneous doping. Thus, in $BaFe_2(As_{1-x}P_x)_2$, the concentration range corresponding to the existence of CT clusters (and hence the position of the superconducting dome) is in the range $0.2 < x \leq 0.5$ [19,20].

In the case of $Ba(Fe_{1-x}Ru_x)_2As_2$, the replacement of the Fe ion by isovalent Ru with a large ionic radius is equivalent to an increase in the negative charge near 4 surrounding As ions, which reduces the gap for electron transfer from these As ions to neighboring Fe ions (Fig. .3e). We will denote such As ions as As', and As ions whose neighbors are two Ru ions, we will denote As". At Ru concentration $x > 0.2$, clusters of $FeAs'_4$ CT plaquettes will be formed, and at larger $x$, $FeAs''_4$ clusters begin to predominate. Therefore, the range of Ru concentrations corresponding to the existence of CT clusters in the phase diagram of $Ba(Fe_{1-x}Ru_xAs)_2$ coincides with the same range of P concentrations in $BaFe_2(As_{1-x}P_x)_2$ (Fig. 3a-d).

Thus, the ranges found for the existence of CT clusters in compounds of isovalently doped ferropnictides $BaFe_2(As_{1-x}P_x)_2$ and $Ba(Fe_{1-x}Ru_x)_2As_2$ coincide with the positions of superconducting domes in their phase diagrams [19–21].

As is known, isovalently doped superconducting compounds have been found only in the series of ferropnictides. Is the same doping possible in cuprates? Due to the symmetry of the basal plane, this type of doping can only be achieved by replacing Cu with an ion of a larger radius (for example, Ag). However, due to the large difference between the radii of Cu and Ag, it is impossible to carry out substitution in the required proportion (~50%). At the same time, it is known that the substitution of 5% Cu for Ag in nonsuperconducting $La_2CuO_4$ actually results in the formation of a superconducting phase with Tc = 28 K [22].

**Heitler-london centres**

The fact that for both classes of compounds all information concerning the position of superconducting domes on their phase diagrams and other "magic" values of dopant concentrations can be obtained based on knowledge of only the crystal structure and type of dopant, means that in CT clusters formed in these materials by doping, a special mechanism of superconductivity operates, which is insensitive to the details of the electronic structure. Therefore, let us consider what properties of the CT cluster may turn out to be important for such a pairing mechanism.

First of all, let us consider the most important property of the percolation CT cluster, which determines the unique properties of cuprates and ferropnictides. Let us take two CT plaquettes belonging to the CT cluster and centred on the nearest same ions Such a pair of plaquettes is a solid-state analog of a hydrogen molecule and can be considered as a Heitler–London (HL) centre [12]

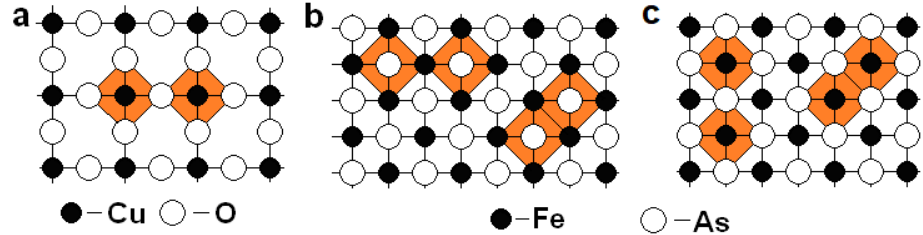

FIG. 4. Various types of HL centres in cuprates and ferropnictides. (a), in cuprates, only one type of HL centres can be formed; (b), in ferropnictides, two types of HL centres are possible, where CT plaquettes have one or two common ions.

Figure 4 presents various types of HL centres in cuprates and ferropnictides. In the case of cuprates, there is only one possibility of the formation of such a centre (Fig. 4a), while in ferropnictides, due to the different structure of the basal plane, two types of HL centres are possible, formed from CT plaquettes $AsFe_4$ (Fig. 4b) or $FeAs_4$ (Fig. 4c). In this case, for each type of HL centres, two different configurations are possible, in which CT plaquettes have one or two common ions. Such HL centres have one remarkable property. Since the transition of an electron (hole) to the central ion in these CT plaquettes requires the presence of a hole (electron) on the outer ions of the plaquette, then on such a hydrogen-like HL centre, two electrons (two holes) on the central ions will form a bound state with two holes (electrons) on the orbitals of external ions due to the possibility of a pair of singlet holes (electrons) being simultaneously between the central ions and being attracted to two electrons (holes) located on these ions. The estimate of the binding energy in this case gives a value of ~0.2 eV [23]. Thus, a CT cluster is a network of CT plaquettes in which each pair of neighboring CT plaquettes is an HL centre, i.e. the CT cluster is simultaneously a cluster of HL centres.

As shown in [12, 13], the range of dopant concentrations corresponding to the superconducting dome coincides with the interval where the existence of CT clusters, which form the superconducting phase, is possible. These clusters can be percolation (optimal phase) or represent a Josephson medium, where these clusters are immersed in a non-superconducting phase (the insulator is in underdoped cuprates, the AFM metal is in underdoped ferropnictides, and the metal is in overdoped cuprates and ferropnictides). For the formation of a CT plaquette, it is necessary that a localized excess charge of a certain magnitude and sign, coming from the nearest dopants, be located near its cation or anion. The value of $T_c$ in a superconducting cluster (percolation or Josephson coupled) is determined by the concentration of CT plaquettes. In the overdoping range, the number of CT plaquettes decreases with concentration due to an increase in the number of plaquettes exposed to several dopants, which leads to vanishing of the /delta_ct in them and a gradual transition to the metal phase. Therefore, the range of existence of superconductivity in the phase diagram has the shape of a dome, where $T_c$ decreases at concentrations below and above the optimal doping range.

Naturally, the formation of large percolation clusters of HL centres, as well as large CT clusters, is possible only with an ordered arrangement of trion complexes (or dopants). As was shown in [13], this is facilitated by their ordering into lattices with a symmetry that retains the main symmetry elements of the matrix crystal.

**Free carrier generation in cuprates and ferropnictides under heterovalent and isovalent doping**

Let us now consider the electronic structure of a CT cluster formed in the $CuO_2$ or FeAs plane by local heterovalent or isovalent doping. As noted, the anomalies in the superconducting characteristics of cuprates and ferropnictides observed at dopant concentrations within superconducting dome, as well as the positions of the domes in phase diagrams, can be understood and calculated with high accuracy basing only on the knowledge of the crystal structure and the type of dopant. Hence, we can conclude that these properties do not depend on the details of the electronic structure in the vicinity of $E_F$, but are determined by other parameters of the band structure common to cuprates and ferropnictides.

Figure 1 c,d shows simplified energy diagrams of the parent phases of these compounds. In the CT cluster, doping reduces the gap $\Delta_{ct}$ to $\Delta_{ct}^* < E_{ex}$ (fig. 5 a,b). As a result, some electrons from the p-band ($O^{2-}$, $As^{3-}$) and holes from the d-band ($Cu^+$, $Fe^{2+}$) pass into the excitonic-bound states $d^+(p^-)$ and $p^-(d^+)$ (Fig. 5c). Here $d^+(p^-)$ is the state of a d-electron on Cu or Fe, in the same plaquette with which there is a p-hole on O or As, respectively, and similarly $p^-(d^+)$ is the state of a p-hole on O or As, in one plaquette with which there is a d-electron on Cu or Fe, respectively. Such a transformation of the band structure of the CT cluster corresponds to its transition to the excitonic insulator state, where two-particle transitions to and from become possible between two one-particle states (p electron + d hole) on the one hand, and an excitonic two-particle state (d electron + p hole) on the other hand. Thus, in both cuprates and ferropnictides, in a certain range of dopant concentrations, doping produces clusters of the CT-excitonic insulator in the $CuO_2$ or FeAs planes. The sizes of these clusters depend on the concentration, which determines their ability to be ordered into lattices with a various symmetry [13]. Note that the transition of the system to the state of an excitonic insulator justifies the simplified description of the band structure, since such a transition couples single-particle states and forms a gap at the Fermi level.

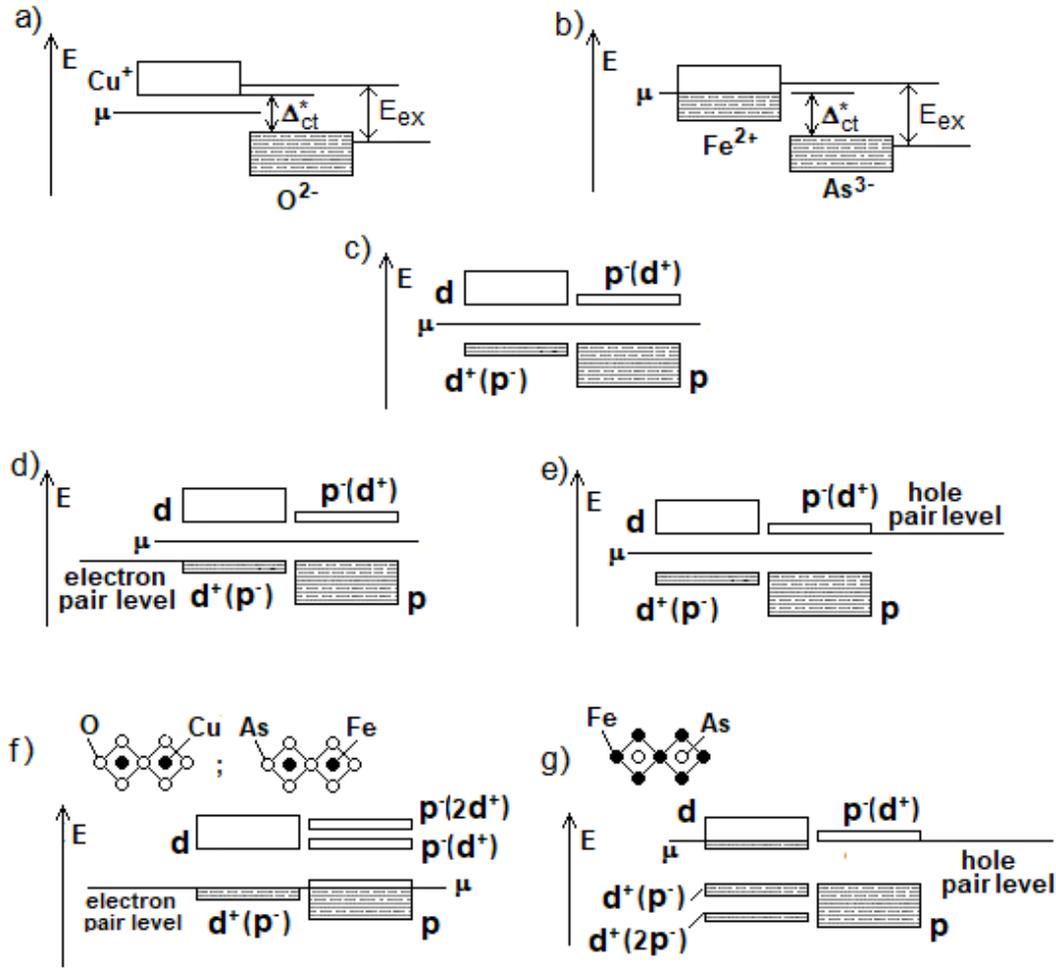

FIG. 5. (a,b), simplified picture of the band structure of doped cuprates (a) and ferropnictides (b), in the CT cluster, doping reduces the gap $\Delta_{ct}$ to $\Delta^*_{ct}<E_{ex}$. As a result, part of the electrons from the p band ($O^{2-}$, $As^{3-}$) and holes from the d band ($Cu^+$, $Fe^{2+}$) pass into the exciton-bound states $d^+(p^-)$ and $p^-(d^+)$. Here $d^+(p^-)$ is the state of a d-electron on Cu or Fe, in the same plaquette with which there is a p-hole on O or As, respectively, and, similarly, $p^-(d^+)$ is the state of a p-hole on O or As, in one plaquette with which there is a d-electron on Cu or Fe, respectively. This corresponds to the transition of the CT cluster to the state of an excitonic insulator; (d), electron pair level of HL centre is the level of coupled pairs of $d^+(p^-)$ electrons localized on the central cations of two neighboring plaquettes; (e), hole pair level of HL centre is the level of coupled pairs of $p^-(d^+)$ holes localized on the central anions of two neighboring plaquettes; (f), two electrons on the central cations of neighboring plaquettes and one hole on the outer anions form a bound state (ionized HL center). Here $p^-(2d^+)$ is the state of hole on the outer O or As anions, which binds two $d^+(p^-)$ electrons on the electron pair level of neighboring Cu or Fe cations. The free holes that arise from the ionization of HL centers form a subband of free hole carriers at the chemical potential level; (g), two holes in the central anions of neighboring plaquettes and one electron on the outer cations form a bound state (ionized HL center). Here $d^+(2p^-)$ is the state of the electron on the Fe cations, which binds two $p^-(d^+)$ holes on hole pair level of neighboring As anions. The free electrons that arise from the ionization of HL-centers form an subband of free electron carriers at the chemical potential level.

It should be especially noted that at a temperature lower than the localization temperature of doped carriers $T_{loc}$ the doping does not in itself give additional free carriers, but, on the contrary, leads to the transition of the CT cluster to the state of an excitonic insulator. Nevertheless, as is known, free carriers are present in the system, although their concentration, according to Hall measurements, decreases with temperature. What is the mechanism of free carrier generation in this model?

The point, in our opinion, is the presence of HL centres, which, as we will see, play the role of acceptors (donors) in this case, and only for this reason doping is accompanied by the generation of additional free carriers. Let's consider this mechanism in more detail. As we noted above, two CT plaquettes of the CT cluster centreed on the nearest cations or anions represent a HL centre where two electrons (two holes) on the central ions form a bound state with two holes (electrons) on the orbital of outer ions. In this case, bound pairs of electrons $d^+(p^-)$ on the central ions of two adjacent plaquettes occupy an electron pair level (Fig. 5d). Similarly, bound pairs of holes $p^-(d^+)$ on the central ions of two adjacent plaquettes occupy a hole pair level (Fig. 5e).

Continuing the analogy with hydrogen molecule, we note that, in addition to $H_2$ molecule, there is also a bound state of two protons and one electron, namely, $H_2^+$ ion. In our case, this corresponds to an ionized HL centre, on which two electrons in the centres of neighboring plaquettes and one hole on the surrounding anions, or two holes in the centres of neighboring plaquettes and one electron on the surrounding cations, form a bound state. On fig. 5 (f,g): $p^-(2d^+)$ are the states of holes on the O or As anions, which bind, respectively, the $d^+(p^-)$ electron pairs on the neighboring Cu and As cations and occupy the corresponding electron pair levels. The free holes arising at the ionization of HL-centres form a hole subband of free carriers at the level of the chemical potential µ. Similarly, $d^+(2p^-)$ is the state of an electron on Fe cation that binds a $p^-(d^+)$ hole pair on neighboring As anions and occupy a hole pair level. The free electrons arising at the ionization of HL-centres form the electron subband of free carriers. at the chemical potential level.

Thus, in the case of cuprates and ferropnictides, the ionization of filled HL centres is the mechanism of generation of free carriers in the system. The concentration of free carriers $n$ will be determined from the condition of equality of chemical potentials for electron (hole) pairs at the pair level and holes (electrons) in subbands [24]. For doping levels that are not too high, the concentration of additional free carriers arising due to the population of the pair level will depend on temperature as $n \propto T$ [15]. We note that just such a dependence is observed in cuprates, where the existence of carriers of only such a nature is possible [25, 26].

As HL centres centred on Cu cations (Fig. 4 a) are possible in cuprates, carriers in them at any (electron or hole) type of doping (Fig. 2 a–c) will always be holes (Fig. 5 f). This explains the transition to the hole type of conduction observed in electron-doped cuprates as the temperature decreases below the localization temperature of doped carriers [27–29].

Different cases are possible in ferropnictides, depending on which plaquettes (AsFe$_4$ or FeAs$_4$) form the cluster of HL centers. In the former case (Fig. 5 f), holes will form; in the latter (Fig. 5 g), electrons. It is obvious that the sign of the emerging carriers will not depend on the configuration of the HL centre, i.e. on whether the CT plaquettes have one or two common ions in the HL centre (Fig. 4 b,c). Since dopants form $AsFe_4$ CT clusters in almost all known ferropnictides, free electrons are generated in them. And only in $Ba(Fe_{1-x}Ru_xAs)_2$, where dopants form $FeAs_4$ CT clusters, such carriers are holes [2, 30].

To confirm this conclusion, let us consider the hole-doped compound $Ba_{1-x}K_xFe_2As_2$. In this compound, each doped hole is distributed over 4 nearest As ions [12], forming a trion complex that includes four CT $AsFe_4$ plaquettes and, therefore, according to Fig. 5g, additional electron-type carriers should be generated. Indeed, it can be observed experimentally that the contribution of electron carriers becomes noticeable at low temperatures, when doped holes are localized [31–33], and also at high temperatures, when the concentration of free electrons generated due to ionization of HL centres becomes sufficiently high [34]. In the first case, this contribution manifests itself in the suppression of the growth (and even decrease) of the Hall constant at $T<100$ K [31–33], and in the second case, in the change of sign of the transverse thermal conductivity to negative at $T>150K$ [34].

**On a possible mechanism of superconducting pairing**

Since the states of electrons in a CT cluster can be considered as a superposition of band and exciton states, it is natural to associate the pairing mechanism with the formation of a virtual biexciton bound state at the HL centre. In this case, the basal planes of doped cuprates and ferropnictides can be considered as another type of structures in addition to one-dimensional Little chains [35] and Ginsburg "sandwiches" [36], where the excitonic mechanism of superconductivity can be realized.

This interaction will be effective if the electronic (hole) pair levels of HL centres are not occupied by real electrons (holes), and therefore are vacant to scattering processes with the formation of a virtual biexciton state. Since the process of populating HL centres with real electrons (holes) is accompanied by the generation of free carriers, the efficiency of the interelectronic interaction due to their scattering on HL centres will increase simultaneously with a decrease in the concentration of free carriers n, which depends on temperature as $n \propto T$ [12,23].

According to the above said, the noncoherent transport via a CT cluster at $T>0$ is performed by carriers emerging at the occupation of paired HL centres. At $T\rightarrow0$ their concentration $n\rightarrow0$, and at $T=0$ no noncoherent transport will be possible due to the absence of free carriers. At the same time, such a system in which an electron and a hole are always present in a CT plaquette permits a coherent transport, when all carriers of the same sign move coherently, as a single whole, e.g., as a superconducting condensate. The latter is possible at a superconducting pairing which emerges due to the formation of a bound state of two electrons (holes) scattered into paired states of an HL centre and two holes (electrons) remaining automatically on the surrounding ions of plaquette.

**Specific features of LiFeAs**

As noted earlier, in the majority of doped HTSC cuprates and ferropnictides, superconducting clusters of HL centres do not fill the entire base plane, but form a Josephson medium Only in some compounds where doped carriers are localized outside the basal plane, this entire plane can be a percolation cluster of HL centres.

A special place among HTS compounds is "111" type ferropnictide LiFeAs, which is a superconductor in the absence of doping. In accordance with the proposed model, this means that in this compound (if the possibility of self-doping is excluded), the condition $\Delta_{ct}<E_{ex}$ is initially satisfied, as a result of which the entire base plane can be considered as a single CT cluster.

In support of this statement, we can cite the results of experiments on heterovalent doping of another compound of the 111 type - NaFeAs(Co), which we will compare with the results of similar experiments on a compound of the 122 type - $Ba(Fe_{1-x}Ni_x)_2As_2$. In contrast to LiFeAs, in NaFeAs superconductivity (filamentary type) in the undoped state is observed sporadically, with $T_c$ in the range from 0 to 10 K [36-39]. However, at P > 4 GPa, bulk superconductivity is already observed for all samples with $T_c$ above 30 K. Similarly, bulk superconductivity is consistently observed in Co-doped NaFeAs at x>0.01 with a maximum $T_c$ ~20 K in the range of 0.02<x<0.04 [36]. At the same time, doping with Co in LiFeAs only monotonically lowers $T_c$ [40]. Based on these results, we believe that in undoped NaFeAs the gap $\Delta_{ct}\approx E_{ex}$, i.e. slightly exceeds the value of $\Delta_{ct}$ in LiFeAs. Therefore, small changes in the lattice parameters or deviations from stoichiometry in NaFeAs can locally change the ratio $\Delta_{ct}\approx E_{ex}$ in one direction or another, leading to the appearance or disappearance of filamentary superconductivity. Within the framework of the model under consideration, doped electrons reduce the gap $\Delta_{ct}$ in plaquettes at the outer boundary of their localization area to the value of $0<\Delta_{ct}^{*}<E_{ex}$. In this case, if $\Delta_{ct}$ is reduced throughout the crystal, then the doped carrier localization area, on the boundary of which the condition $0<\Delta_{ct}^{*}<E_{ex}$ is satisfied, will expand. In this case, inside the carrier localization area $\Delta^*=0$. Therefore, in the initially superconducting LiFeAs, where $\Delta_{ct}<E_{ex}$ takes place, we deal with a gradual decrease in superconductivity upon doping and its complete disappearance at some x, when the percolation threshold over the doped carrier localization areas is reached. In the case of initially non-superconducting NaFeAs, where, according to the assumption, $\Delta_{ct} \sim E_{ex}$, doping with Co reduces the value of $\Delta_0$ in plaquettes at the outer boundary of the doped carrier localization area to $0 < \Delta_{ct}^* < E_{ex}$ and thus forms superconducting phase clusters from them [13]. To verify this statement, let us compare the phase diagrams of NaFe$_{1-x}$Co$_x$As[36,37,41] and Ba(Fe$_{1-x}$Ni$_x$)$_2$As$_2$[42]. It is easy to see that the positions of the superconducting domes on them practically coincide despite the fact that a Ni dopant brings two extra electrons to FeAs plane, in contrast to a Co dopant that introduces one extra electron [43].

According to the model under consideration [13], the positions of domes and features of the phase diagrams of cuprates and ferropnictides with heterovalent doping are entirely determined by the geometry of trion complexes (Fig. 2). Based on this, the coincidence of the phase diagrams of NaFe$_{1-x}$Co$_x$As and Ba(Fe$_{1-x}$Ni$_x$)$_2$As$_2$ means that the trion complexes formed in them are the same and have the form shown in Fig. 2f. To make sure that the similarity of the phase diagrams of NaFe$_{1-x}$Co$_x$As and Ba(Fe$_{1-x}$Ni$_x$)$_2$As$_2$ is due to the same geometry of the trion complexes formed in them, let us pay attention to the presence of features in both phase diagrams at two identical values $x=0.027$ and $0.032$ [37 (Fig.4), 45 (Fig.3b)]. The appearance of features at given values of $x$, corresponding to $x=1/36$ and $x=1/32$, is due to the formation of large ordered clusters from trion complexes with a given geometry, repeating the symmetry of the matrix [13]. For Ba(Fe$_{1-x}$Ni$_x$)$_2$As$_2$, the geometry of the trion complex (Fig. 2f) is determined by the fact that 2 doped electrons from Ni completely suppress $\Delta$ for the nearest belt of FeAs$_4$ plaquettes (in contrast to Ba(Fe$_{1-x}$Co$_x$)$_2$As$_2$ in Fig. 2e) and their localization area expands until the excess charge on the Fe ions of the next coordination sphere becomes equal to $e/4$.

The similarity of the phase diagrams of NaFe$_{1-x}$Co$_x$As and Ba(Fe$_{1-x}$Ni$_x$)$_2$As$_2$ means that in NaFe$_{1-x}$Co$_x$As one doped electron from the Co ion completely suppresses $\Delta$ for the nearest FeAs$_4$ belt of plaquettes, and the area of its localization expands until the excess charge on the Fe ions from the next coordination sphere will not become equal to e/8, which will correspond to the condition $0 < \Delta_{ct}^* < E_{ex}$. It does mean, the value of $\Delta_0$ in undoped NaFeAs is much smaller than in other classes of heterovalently doped ferropnictides. This is all the more true in LiFeAs, where $\Delta_0$ is smaller than in NaFeAs.

That confirms our assumption that in LiFeAs the condition $\Delta_{ct} < E_{ex}$ is realized without doping, i.e. the entire base plane can be considered as a single CT cluster.

As previously noted, the formation of CT clusters from AsFe$_4$ plaquettes is accompanied by the formation of additional electron carriers, while the formation of CT clusters from FeAs$_4$ plaquettes leads to the appearance of hole carriers (Fig. 5). This raises the question of the type of carriers in LiFeAs compound where the basal plane can be considered as filled with small CT clusters of various types of CT plaquettes either AsFe4 or FeAs4. Herewith, carriers of both types should be generated. The small size of the clusters prevents the formation of a large coherent cluster in which moving carriers scatter into pair states (either electron or hole). This can explain the relatively low $T_c$ = 17 K in this compound and the observed two-phase nature within the coherence area [45].

**Conclusions**

Thus, in this work, we propose a unified mechanism of the transformation of the electronic structure of cuprates and ferropnictides upon heterovalent and isovalent doping. In this representation, in undoped cuprates and ferropnictides, which initially have different electronic structures (Mott insulator and AFM semimetal), local doping forms hydrogen-like CT plaquettes, from which, in a certain doping range, percolation CT clusters are formed that have the electronic structure of an excitonic insulator.

The most important property of a CT cluster is that each pair of adjacent CT plaquettes in it is an HL center resembling an intracrystalline hydrogen molecule, on which two electrons and two holes can form a bound state. In such a CT cluster, free carriers are generated as a result of partial ionization of filled HL centers. The sign of the generated free carriers in the general case does not coincide with the sign of the doped carriers, but is determined by the type of CT plaquettes that make up the HL center.

As was shown in [10, 12], most of the anomalies in the superconducting characteristics of cuprates and ferropnictides observed at dopant concentrations within the superconducting dome, as well as the very position of the domes in the phase diagrams, do not require knowledge of the details of their electronic structure for their explanation, but can be understood and calculated with great accuracy within a simple model, based only on the knowledge of the crystal structure and the type of dopant. The fact that the positions of the superconducting domes in the phase diagrams coincide with the intervals of existence of percolation clusters of CT plaquettes suggests that high-temperature superconductivity takes place in a CT cluster where band electrons can interact with biexciton states of HL centers. This makes it possible to consider HTSC as a medium where the realization of the exciton mechanism of superconductivity is possible.

**CRediT authorship contribution statement**

**K. Mitsen**:. Conceptualization, Model construction, Writing – original draft. **O. Ivanenko**. Development of this work, Supervision, Writing – review & editing.

**Declaration of competing interest**

The authors declare that they have no known competing financial interests or personal relationships that could have appeared to influence the work reported in this paper.